\documentclass[aps,prb,reprint,showpacs]{revtex4-1}

\usepackage{graphicx}

\begin{document}


\title{Ferroelectric properties of RbNbO$_3$ and RbTaO$_3$}


\author{A. I. Lebedev}
\email[]{swan@scon155.phys.msu.ru}
\affiliation{Physics Department, Moscow State University, Moscow, 119991 Russia}


\date{\today}

\begin{abstract}
Phonon spectra of cubic rubidium niobate and rubidium tantalate with the
perovskite structure are calculated from first principles within the density
functional theory. Based on the analysis of unstable modes in the phonon
spectra, the structures of possible distorted phases are determined, their
energies are calculated, and it is shown that $R3m$ is the ground-state structure
of RbNbO$_3$. In RbTaO$_3$, the ferroelectric instability is suppressed by
zero-point lattice vibrations. For ferroelectric phases of RbNbO$_3$,
spontaneous polarization, piezoelectric, nonlinear optical, electro-optical,
and other properties as well as the energy band gap in the LDA and $GW$
approximations are calculated. The properties of the rhombohedral RbNbO$_3$
are compared with those of rhombohedral KNbO$_3$, LiNbO$_3$, and BaTiO$_3$.

DOI: 10.1134/S1063783415020237
\end{abstract}


\maketitle

The possibility of ferroelectricity in rubidium niobate and rubidium tantalate
with the perovskite structure was discussed by Smolenskii and
Kozhevnikova~\cite{DoklAkadNauk.76.519} and then by Megaw~\cite{ActaCryst.5.739}
in the early 1950s. In Ref.~\onlinecite{DoklAkadNauk.76.519}, the authors
referred to unpublished data by V.G. Prokhvatilov who detected the tetragonal
RbTaO$_3$ phase with $a = 3.92$~{\AA}, $c = 4.51$~{\AA} exhibiting a phase
transition near 520~K; in Ref.~\onlinecite{ActaCryst.5.739} these data were
simply cited. However, further studies have shown that, unlike lithium, sodium,
and potassium niobates, RbNbO$_3$ and RbTaO$_3$ crystallize in individual
crystal structures with triclinic $P{\bar 1}$ symmetry for RbNbO$_3$ and
monoclinic $C2/m$ one for RbTaO$_3$~\cite{JLessCommonMetals.76.299,
AngewandteChem.90.387,ZAnorgAllgChem.464.240} when prepared at atmospheric
pressure. To obtain these materials with the perovskite structure, they should
be prepared at high pressures (65--90~kbar).~\cite{Kafalas72}  Due to the
difficulties in synthesis of RbNbO$_3$ and RbTaO$_3$ with the perovskite
structure, the properties of these crystals have been studied very little.

The phase diagrams of Rb$_2$O--Nb$_2$O$_5$ and Rb$_2$O--Ta$_2$O$_5$ systems
were studied in Ref.~\onlinecite{JPhysChem.64.748,MaterResBull.11.299}.
RbNbO$_3$ is formed by the peritectic reaction and decomposes above
964$^\circ$C.\cite{JPhysChem.64.748}  RbTaO$_3$ decomposes above 600$^\circ$C
probably due to the peritectic reaction too.~\cite{MaterResBull.11.299}
Rubidium-containing ferroelectric materials in the
BaNb$_2$O$_6$--NaNbO$_3$--RbNbO$_3$ system with the tungsten-bronze structure
have high electro-optical properties that substantially exceed those of lithium
niobate.~\cite{JElectrochemSoc.116.1555,USPatent.3640865}  In
Ref.~\onlinecite{EnergyEnvironSci.5.9034}, the possibility of using rubidium
niobate and rubidium tantalate for photoelectrochemical decomposition of water
was discussed. In Ref.~\onlinecite{InorgChem.46.4787}, it was proposed to use
the delamination of RbTaO$_3$ structure to produce porous TaO$_3$ nanomembranes
with pore sizes of 1.3$\times$0.6 and 1.1$\times$1.1~{\AA}, which can be used
for selective filtration of lithium ions.

The lack of knowledge on the properties of compounds under consideration
appears, in particular, in contradictory data on the ferroelectric properties
of RbTaO$_3$. For example, the existence of the phase transition at 520~K in
the tetragonal phase was reported in Ref.~\onlinecite{DoklAkadNauk.76.519},
whereas the data of Ref.~\onlinecite{Kafalas72} showed that RbTaO$_3$ prepared
at high pressure has the cubic perovskite (or close to it) structure. At 300~K,
the structure of RbNbO$_3$ is similar to that of the orthorhombic BaTiO$_3$,
and the data of differential thermal analysis indicate phase transitions in it
at 15, 155, and 300$^\circ$C.~\cite{Kafalas72}

In this work, the equilibrium structures of RbNbO$_3$ and RbTaO$_3$ were
determined from first-principles calculations, and spontaneous polarization,
dielectric constant, piezoelectric and elastic moduli, nonlinear optical and
electro-optical properties as well as the energy band gaps in the LDA and $GW$
approximations were calculated for these crystals.

\begin{table}
\caption{\label{table1}Calculated lattice parameters and atomic coordinates in
RbNbO$_3$ structures.}
\begin{ruledtabular}
\begin{tabular}{ccccc}
Atom  & Position & $x$        & $y$     & $z$ \\
\hline
\multicolumn{5}{c}{\rule{0pt}{4.2mm}Phase $P{\bar 1}$} \\
\multicolumn{5}{c}{$a = 5.0816$, $b = 8.3047$, $c = 8.7916$~{\AA},} \\
\multicolumn{5}{c}{$\alpha = 114.0625$, $\beta = 93.3891$, $\gamma = 95.1160$$^\circ$} \\
\hline
Rb2   & $1a$  & +0.00000 & +0.00000 & +0.00000 \\
Rb1   & $1b$  & +0.00000 & +0.00000 & +0.50000 \\
Rb3   & $2i$  & +0.41251 & +0.70257 & +0.09488 \\
Nb1   & $2i$  & +0.49674 & +0.28138 & +0.35602 \\
Nb2   & $2i$  & +0.02746 & +0.51037 & +0.30988 \\
O1    & $2i$  & +0.10125 & +0.39078 & +0.82160 \\
O2    & $2i$  & +0.23664 & +0.42747 & +0.50994 \\
O3    & $2i$  & +0.28650 & +0.71922 & +0.45293 \\
O4    & $2i$  & +0.29777 & +0.37205 & +0.19910 \\
O5    & $2i$  & +0.33951 & +0.05579 & +0.27137 \\
O6    & $2i$  & +0.78420 & +0.27571 & +0.22313 \\
\hline
\multicolumn{5}{c}{Phase $Pm3m$} \\
\multicolumn{5}{c}{$a = 4.0291$~{\AA}} \\
\hline
Rb    & $1a$  & +0.00000 & +0.00000 & +0.00000 \\
Nb    & $1b$  & +0.50000 & +0.50000 & +0.50000 \\
O     & $3c$  & +0.00000 & +0.50000 & +0.50000 \\
\hline
\multicolumn{5}{c}{Phase $P4mm$} \\
\multicolumn{5}{c}{$a = 4.0037$, $c = 4.1592$~{\AA}} \\
\hline
Rb    & $1a$  & +0.00000 & +0.00000 & $-$0.00336 \\
Nb    & $1b$  & +0.50000 & +0.50000 &   +0.51818 \\
O1    & $2c$  & +0.50000 & +0.00000 &   +0.47446 \\
O2    & $1b$  & +0.50000 & +0.50000 & $-$0.03654 \\
\hline
\multicolumn{5}{c}{Phase $Amm2$} \\
\multicolumn{5}{c}{$a = 3.9928$, $b = 5.7742$, $c = 5.7960$~{\AA}} \\
\hline
Rb    & $2a$  & +0.00000 & +0.00000 & $-$0.00341 \\
Nb    & $2b$  & +0.50000 & +0.00000 &   +0.51447 \\
O1    & $4e$  & +0.50000 & +0.25496 &   +0.22842 \\
O2    & $2a$  & +0.00000 & +0.00000 &   +0.47735 \\
\hline
\multicolumn{5}{c}{Phase $R3m$} \\
\multicolumn{5}{c}{$a = 4.0571$~{\AA}, $\alpha = 89.8945^\circ$} \\
\hline
Rb    & $1a$  & $-$0.00308 & $-$0.00308 & $-$0.00308 \\
Nb    & $1a$  &   +0.51193 &   +0.51193 &   +0.51193 \\
O     & $3b$  & $-$0.02115 &   +0.48415 &   +0.48415 \\
\end{tabular}
\end{ruledtabular}
\end{table}

\begin{table}
\caption{\label{table2}Calculated lattice parameters and atomic coordinates
in RbTaO$_3$ structures.}
\begin{ruledtabular}
\begin{tabular}{ccccc}
Atom & Position & $x$        & $y$     & $z$ \\
\hline
\multicolumn{5}{c}{Phase $C2/m$} \\
\multicolumn{5}{c}{$a = b = 6.3396$, $c = 8.0171$,} \\
\multicolumn{5}{c}{$\alpha = 86.1031$, $\beta = 93.8969$, $\gamma = 96.8997^\circ$} \\
\hline
Rb1  & $4i$ & +0.16000 & $-$0.16000 & +0.73758 \\
Rb2  & $4g$ & +0.26494 &   +0.26494 & +0.00000 \\
Ta1  & $4h$ & +0.31104 &   +0.31104 & +0.50000 \\
Ta2  & $4i$ & +0.23924 & $-$0.23924 & +0.30214 \\
O1   & $8j$ & +0.27303 &   +0.04639 & +0.39643 \\
O2   & $8j$ & +0.54749 & $-$0.21209 & +0.28704 \\
O3   & $4i$ & +0.16752 & $-$0.16752 & +0.08653 \\
O4   & $4i$ & +0.37705 & $-$0.37705 & +0.55235 \\
\hline
\multicolumn{5}{c}{Phase $Pm3m$} \\
\multicolumn{5}{c}{$a = 3.9846$~{\AA}} \\
\hline
Rb   & $1a$ & +0.00000 & +0.00000 & +0.00000 \\
Ta   & $1b$ & +0.50000 & +0.50000 & +0.50000 \\
O    & $3c$ & +0.00000 & +0.50000 & +0.50000 \\
\end{tabular}
\end{ruledtabular}
\end{table}

The first-principles calculations were performed within the density functional
theory using the \texttt{ABINIT} software.~\cite{abinit2} The
exchange--correlation interaction was described in the local density
approximation (LDA). The optimized norm-conserving pseudopotentials for Nb, Ta,
and O atoms used in these calculations were taken from
Ref.~\onlinecite{PhysSolidState.52.1448}.  The non-relativistic pseudopotential
for the Rb atom (electronic configuration $4s^24p^65s^0$) was constructed
according to the scheme of Ref.~\onlinecite{PhysRevB.41.1227} using the
\texttt{OPIUM} program~\cite{opium} with the following parameters: $r_s = 1.68$,
$r_p = 1.72$, $r_d = 1.68$, $q_s = 7.07$, $q_p = 7.27$, $q_d = 7.07$,
$r_{\rm min} = 0.01$, $r_{\rm max} = 1.52$, and $V_{\rm loc} = 1.58$~a.u. (for
notations, see Ref.~\onlinecite{PhysSolidState.51.362}). The testing of the Rb
pseudopotential on the $P\bar{1}$ phase of RbNbO$_3$ and the $C2/m$ phase of
RbTaO$_3$, which are stable at atmospheric pressure, showed its sufficiently
high quality: the calculated lattice parameters and atomic coordinates in these
phases (see Tables~\ref{table1} and \ref{table2}) are in good agreement with the
experimental data;~\cite{JLessCommonMetals.76.299,ZAnorgAllgChem.464.240} small
underestimates of the calculated lattice parameters are characteristic of the
LDA approximation used in this work.

The lattice parameters and equilibrium atomic positions in the unit cells were
determined from the condition when the residual forces acting on the atoms were
below 5$\cdot$10$^{-6}$~Ha/Bohr (0.25~meV/{\AA}) in the self-consistent
calculation of the total energy with an accuracy better than 10$^{-10}$~Ha.
The maximum energy of plane waves was 30~Ha for RbNbO$_3$ and 40~Ha for
RbTaO$_3$. Integration over the Brillouin zone was performed using a
8$\times$8$\times$8 Monkhorst--Pack mesh. The spontaneous polarization in
ferroelectric phases was calculated by the Berry phase method. The phonon
spectra, dielectric constants, piezoelectric and elastic moduli were calculated
within the density-functional perturbation theory similarly to
Ref.~\onlinecite{PhysSolidState.51.362}.  Nonlinear optical and electro-optical
properties were calculated using the technique described in
Ref.~\onlinecite{PhysRevB.71.125107}. All physical properties presented in this
work were calculated for the theoretical lattice parameter.

\begin{figure}
\centering
\includegraphics{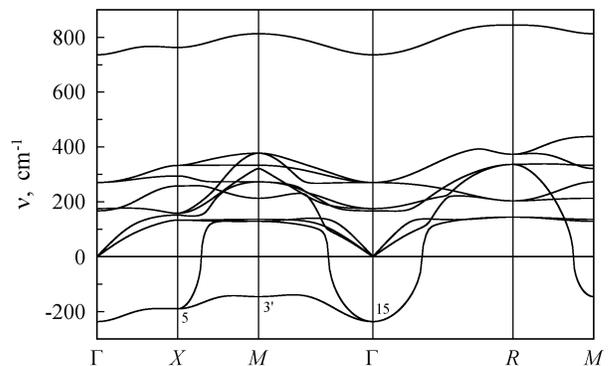}
\caption{Phonon spectrum of RbNbO$_3$ in the cubic $Pm3m$ phase. Labels near
curves indicate the symmetry of unstable modes.}
\label{fig1}
\end{figure}

The phonon spectrum of RbNbO$_3$ in the cubic $Pm3m$ phase is shown in Fig.~1.
This spectrum contains a band of unstable modes characteristic of ferroelectric
chain instability which was first observed in KNbO$_3$.~\cite{PhysRevLett.74.4067}
At the center of the Brillouin zone, this mode has the $\Gamma_{15}$ symmetry,
is triply degenerate, and describes the ferroelectric distortion of structure.
The structures appearing upon condensation of the $X_5$ and $M_3'$ modes are
characterized by antiparallel orientation of polarization in neighboring
...--O--Nb--O--... chains.

\begin{table}
\caption{\label{table3}Relative energies of low-symmetry RbNbO$_3$ phases formed
from the cubic perovskite phase upon condensation of unstable phonons, phases
with $6H$, $4H$, $9R$, and $2H$ structures, and the $P{\bar 1}$ phase prepared
at atmospheric pressure (the most stable phase energy is in boldface).}
\begin{ruledtabular}
\begin{tabular}{ccc}
Phase         &	Unstable mode &	Energy, meV \\
\hline
$Pm3m$        & ---           & 0 \\
$P4/nmm$      & $M_3'$        & $-$31.3 \\
$Pmma$        & $X_5$         & $-$34.8 \\
$Cmcm$        & $X_5$         & $-$38.7 \\
$P4mm$        & $\Gamma_{15}$ & $-$46.5 \\
$Amm2$        & $\Gamma_{15}$ & $-$57.0 \\
$R3m$         & $\Gamma_{15}$ & \bf{$-$58.6} \\
\hline
\rule{0pt}{4.2mm}%
$P{\bar 1}$         & ---     & +27.4 \\
$P6_3/mmc$ ($6H$)   & ---     & +121.4 \\
$P6_3/mmc$ ($4H$)   & ---     & +334.0 \\
$R{\bar 3}m$ ($9R$) & ---     & +568.1 \\
$P6_3/mmc$ ($2H$)   & ---     & +1752 \\
\end{tabular}
\end{ruledtabular}
\end{table}

The energies of all RbNbO$_3$ phases formed upon condensation of the above
unstable modes are given in Table~\ref{table3}. Among these phases, the $R3m$
phase has the lowest energy. The phonon spectrum calculations for the $R3m$
phase show that the frequencies of all optical phonons at the center of the
Brillouin zone and at high-symmetry points at its boundary are positive; the
determinant and all leading principal minors constructed of elastic moduli
tensor components are also positive. This means that the $R3m$ phase is the
ground-state structure of RbNbO$_3$. The calculated lattice parameters and
atomic coordinates in this phase are given in Table~\ref{table1}. As the same
sequence of phases as in BaTiO$_3$ is supposed in rubidium niobate with the
perovskite structure,~\cite{Kafalas72} the lattice parameters and atomic
coordinates in two other ferroelectric phases are also given in this table.
The lattice parameters calculated for the orthorhombic RbNbO$_3$ are in good
agreement with the experimental data obtained at 300~K ($a = 3.9965$,
$b = 5.8360$, and $c = 5.8698$~{\AA}, Ref.~\onlinecite{Kafalas72}).

\begin{table}
\caption{\label{table4}Relative energies of low-symmetry RbTaO$_3$ phases formed
from the cubic perovskite phase upon condensation of unstable phonons, phases
with $6H$, $4H$, $9R$, and $2H$ structures, and the $C2/m$ phase prepared at
atmospheric pressure (the most stable phase energy is in boldface).}
\begin{ruledtabular}
\begin{tabular}{ccc}
Phase         &	Unstable mode & Energy, meV \\
\hline
$Pm3m$        & ---           & 0 \\
$P4mm$        & $\Gamma_{15}$ & $-$1.80 \\
$Amm2$        & $\Gamma_{15}$ & $-$1.87 \\
$R3m$         & $\Gamma_{15}$ & \bf{$-$1.90} \\
\hline
$C2/m$              & ---     & +38.5 \\
$P6_3/mmc$ ($6H$)   & ---     & +113.3 \\
$P6_3/mmc$ ($4H$)   & ---     & +352.2 \\
$R{\bar 3}m$ ($9R$) & ---     & +589.0 \\
$P6_3/mmc$ ($2H$)   & ---     & +1839 \\
\end{tabular}
\end{ruledtabular}
\end{table}

\begin{figure}
\centering
\includegraphics{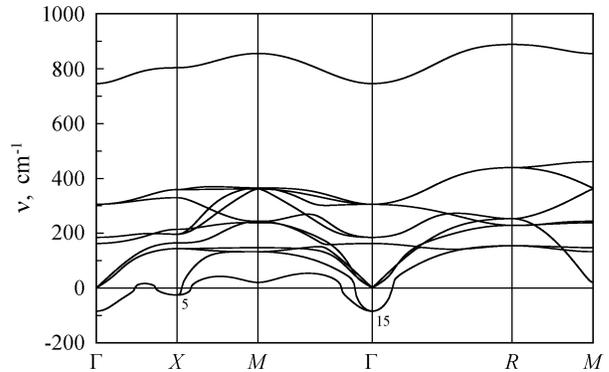}
\caption{Phonon spectrum of RbTaO$_3$ in the cubic $Pm3m$ phase. Labels near
curves indicate the symmetry of unstable modes.}
\label{fig2}
\end{figure}

In RbTaO$_3$, the frequency of unstable $\Gamma_{15}$ phonon in the phonon
spectrum (Fig.~2) and the energy gain resulting from the transition to
ferroelectric phases (Table~\ref{table4}) are rather low; so it is necessary to
additionally test the stability of the ferroelectric distortion with respect
to zero-point lattice vibrations. For this purpose, we used the technique
proposed in Ref.~\onlinecite{PhysSolidState.51.802}. The energy gain resulting
from the transition from the $Pm3m$ phase to the $R3m$ phase is
$E_0 = 1.90$~meV and the unstable phonon frequency at the $\Gamma$~point in
the $Pm3m$ phase is $\nu = 84$~cm$^{-1}$. As the energy ratio $h\nu/E_0 = 5.51$
exceeds the critical value of 2.419 obtained in
Ref.~\onlinecite{PhysSolidState.51.802}, the energy of the lowest vibrational
state in a two-well potential appears above the upper point of the energy
barrier separating the potential wells, and the ferroelectric ordering is
suppressed by zero-point vibrations. Therefore, the only stable phase of
RbTaO$_3$ with the perovskite structure is the cubic phase. The calculated
lattice parameter of this phase is given in Table~\ref{table2}; its value
is in satisfactory agreement with the experimental data ($a = 4.035$~{\AA},
Ref.~\onlinecite{Kafalas72}).

It is known that the formation of phases with hexagonal BaNiO$_3$ (polytype $2H$),
hexagonal BaMnO$_3$ (polytype $4H$), hexagonal BaTiO$_3$ (polytype $6H$), and
rhombohedral BaRuO$_3$ (polytype $9R$) structures is characteristic of
\emph{AB}O$_3$ perovskites with the tolerance factor $t > 1$, and the studied
compounds belong to this class. Our calculations showed that the energies of
these phases for both rubidium compounds are appreciably higher than the cubic
phase energy (Tables~\ref{table3} and \ref{table4}). These results explain why
it was impossible to observe the transition of RbNbO$_3$ to the hexagonal
structure~\cite{Kafalas72} upon heating, by analogy to that occurring in
BaTiO$_3$. The high energies of these phases, in particular, the $2H$ phase,
are probably caused by larger sizes and strong electrostatic repulsion of
Nb$^{5+}$ ions which occupy face-sharing octahedra in these structures.

\begin{table}
\caption{\label{table5}Nonzero components of the piezoelectric tensor $e_{i\nu}$
(C/m$^2$) and tensors of the second-order nonlinear optical susceptibility
$d_{i\nu}$ and the linear electro-optic effect $r_{i\nu}$ (pm/V) in rhombohedral
phases of RbNbO$_3$, KNbO$_3$, LiNbO$_3$, and BaTiO$_3$.}
\begin{ruledtabular}
\begin{tabular}{ccccc}
Coefficient & RbNbO$_3$ & KNbO$_3$ & LiNbO$_3$ & BaTiO$_3$ \\
\hline
$e_{11}$ &   $-$3.0 &  $-$4.2 &  $-$2.4 & $-$4.0 \\
$e_{15}$ &     +4.8 &    +6.8 &    +3.5 &   +7.3 \\
$e_{31}$ &     +2.4 &    +2.3 &    +0.1 &   +3.5 \\
$e_{33}$ &     +2.9 &    +3.1 &    +1.1 &   +5.1 \\
$d_{11}$ &    +12.7 &   +11.9 &    +2.3 &   +4.4 \\
$d_{15}$ &  $-$23.6 & $-$21.9 & $-$11.5 & $-$16.1 \\
$d_{31}$ &  $-$23.6 & $-$21.9 & $-$11.5 & $-$16.1 \\
$d_{33}$ &  $-$29.4 & $-$27.3 & $-$37.4 & $-$31.1 \\
$r_{11}$ &  $-$12.8 & $-$17.7 & $-$5.6  & $-$13.7 \\
$r_{15}$ &    +27.6 &   +39.2 &   +17.1 &   +43.3 \\
$r_{31}$ &    +18.0 &   +23.9 &   +10.1 &   +25.3 \\
$r_{33}$ &    +30.1 &   +40.6 &   +27.3 &   +48.9 \\
\end{tabular}
\end{ruledtabular}
\end{table}

We consider now some properties of ferroelectric RbNbO$_3$. The calculated
polarization in RbNbO$_3$ is 0.46~C/m$^2$ in the $P4mm$ phase and 0.50~C/m$^2$
in the $Amm2$ and $R3m$ phases; these values slightly exceed the calculated
polarization in the same phases of KNbO$_3$ (0.37, 0.42, and 0.42~C/m$^2$,
respectively). The static dielectric tensor in the $R3m$ phase is characterized
by two eigenvalues: $\varepsilon^0_{\parallel} = 21.1$ and
$\varepsilon^0_{\perp} = 35.8$; the optical dielectric tensor eigenvalues are
$\varepsilon^{\infty}_{\parallel} = 5.31$ and
$\varepsilon^{\infty}_{\perp} = 5.91$. In the cubic phase, the elastic moduli
are $C_{11} = 412$, $C_{12} = 84$, and $C_{44} = 102$~GPa; the bulk modulus is
$B = 193.5$~GPa. The nonzero components of the tensors of piezoelectric effect
$e_{i\nu}$, second-order nonlinear optical susceptibility $d_{i\nu}$, and
linear electro-optic (Pockels) effect $r_{i\nu}$ in the $R3m$ phase of rubidium
niobate are compared with the corresponding properties of other rhombohedral
ferroelectrics in Table~\ref{table5}. We can see that the piezoelectric moduli
in rhombohedral RbNbO$_3$ (as well as in its other polar phases) are slightly
lower than in KNbO$_3$. The nonlinear optical coefficients in RbNbO$_3$ exceed
the corresponding values in KNbO$_3$, although the $d_{33}$ value in rubidium
niobate is slightly lower than in lithium niobate. As for the electro-optical
properties, in rhombohedral RbNbO$_3$ they are slightly lower than in KNbO$_3$,
but are notably superior to those of lithium niobate. In the orthorhombic phase
(stable at 300~K), nonlinear optical properties of RbNbO$_3$ are comparable to
those of the same phase of potassium niobate: for example, the $d_{33}$ modulus
is $-$30.8~pm/V in RbNbO$_3$ and $-$30.4~pm/V in KNbO$_3$.

In cubic RbTaO$_3$, the optical dielectric constant is $\varepsilon_{\infty} = 5.58$.
The static dielectric constant can be estimated only in the rhombohedral phase
as $\sim$140. The elastic moduli in cubic rubidium tantalate are $C_{11} = 466$,
$C_{12} = 91.5$, and $C_{44} = 120$~GPa; $B = 216$~GPa. The piezoelectric
moduli, second-order nonlinear optical susceptibility, and electro-optical
coefficients in the cubic phase are zero.

An unexpected result of our calculations is that in both studied compounds the
$P\bar{1}$ and $C2/m$ phases which can be prepared at atmospheric pressure are
metastable. This result is probably caused by an effective lattice contraction
which always exists in the LDA calculations. The fact that the specific volume
of the $Pm3m$ phase is noticeably smaller than that of $P\bar{1}$, $C2/m$,
$P6_3/mmc$, and $R\bar{3}m$ phases suggests that under pressure the cubic
perovskite phase will be the most stable one. To estimate the maximum value of
the actual effective pressure, the lattice parameters and atomic positions in
the $C2/m$ structure of rubidium tantalate were calculated for different
pressures and it was shown that the unit cell volume equal to the experimental
one at 300~K can be obtained at an isotropic pressure of $-$24.7~kbar. At this
pressure, the enthalpy of the $C2/m$ phase becomes lower than that of $Pm3m$ by
$\sim$230~meV, i.e., becomes consistent with the experimental data. At the
above-mentioned negative pressure, the ratio $h\nu/E_0$ determining the stability
of the ferroelectric phase in RbTaO$_3$ becomes equal to 1.90, i.e., it is
slightly lower than the critical value of 2.419. However, if we take into account
that the above negative pressure is obviously overestimated because it includes
the thermal expansion effect, we can suppose that, even taking into account the
systematic error in the LDA lattice parameter determination, rubidium tantalate
will remain cubic up to the lowest temperatures.

The conclusion that RbTaO$_3$ is an incipient ferroelectric in which the
ferroelectric ordering is suppressed by zero-point vibrations agrees with the
data of Ref.~\onlinecite{Kafalas72}, but contradicts the data of
Ref.~\onlinecite{DoklAkadNauk.76.519} in which the phase transition near 520~K
was reported. We suppose that tantalum-enriched phases (in particular, with
the tungsten-bronze structure~\cite{MaterResBull.11.299}) could be formed in
rubidium tantalate samples discussed in Ref.~\onlinecite{DoklAkadNauk.76.519}
because of the low temperature of the peritectic reaction, and this could result
in the observed anomaly.

In Ref.~\onlinecite{EnergyEnvironSci.5.9034}, the possibility of using various
oxides with the perovskite structure, in particular RbNbO$_3$ and RbTaO$_3$, for
development of photoelectrochemical solar cells was discussed. We calculated the
band gap $E_g$ in these compounds both in the LDA approximation and in the $GW$
approximation that takes into account many-body effects (the technique of the
latter calculations was analogous to that used in
Refs.~\onlinecite{PhysSolidState.54.1663,JAlloysComp.580.487,
PhysSolidState.56.1039}). In the LDA approximation, $E^{\rm LDA}_g = 1.275$~eV
in cubic RbNbO$_3$ when the spin--orbit coupling is neglected; in $P4mm$,
$Amm2$, and $R3m$ phases, $E^{\rm LDA}_g$ is 1.314, 1.869, and 2.137~eV,
respectively. In cubic RbTaO$_3$, $E^{\rm LDA}_g = 2.175$~eV when the
spin--orbit coupling is neglected. The valence band extrema in the cubic phase
of both compounds are at the $R$~point of the Brillouin zone, whereas the
conduction band extrema are at the $\Gamma$~point. The calculations using the
technique of Ref.~\onlinecite{PhysSolidState.56.1039} yield the spin--orbit
splitting of the conduction band edge $\Delta_{SO} = 0.111$~eV for RbNbO$_3$
and $\Delta_{SO} = 0.400$~eV for RbTaO$_3$; the spin-orbit splitting of the
valence band edge is absent. After correction for the conduction band edge
shift ($\Delta_{SO}/3$), the LDA values of $E_g$ that take into account the
spin--orbit coupling are 1.238, 1.277, 1.832, 2.100, and 2.042~eV for four
RbNbO$_3$ phases and for cubic RbTaO$_3$, respectively.

In the $GW$ approximation, the band gap when the spin--orbit coupling is
neglected is $E^{GW}_g = 2.403$, 2.616, 3.291, and 3.609~eV, respectively
in cubic, tetragonal, orthorhombic, and rhombohedral RbNbO$_3$ and 3.302~eV
in cubic RbTaO$_3$. If the spin--orbit coupling is taken into account,
these values decrease to 2.366, 2.579, 3.254, 3.572, and 3.169~eV,
respectively. The obtained values of $E_g$ are appreciably smaller than
those calculated in Ref.~\onlinecite{EnergyEnvironSci.5.9034} for cubic
phases (3.4~eV for RbNbO$_3$ and 4.3~eV for RbTaO$_3$).

Some authors who studied rubidium niobate and rubidium tantalate have noticed
their sensitivity to humidity. Evidently, this can be a serious obstacle for
practical applications of these materials. However, we would like to note that
this property is inherent to phases prepared at atmospheric pressure and having
loose structures whose specific volume is 26--28\% larger than that of the
perovskite phase. In Ref.~\onlinecite{MaterResBull.11.299}, it was suggested
that the effect is due to intercalation of water molecules into the loose
structures, rather than to hydrolysis of these compounds. The possibility of
preparing RbTaO$_3$ by hydrothermal synthesis~\cite{JMaterChemA.2.8033} and
the low rate of RbTaO$_3$ ion exchange in HCl during its
delamination~\cite{InorgChem.46.4787} support this idea. This suggests that
the considered compounds with the perovskite structure can be quite stable to
humidity.

In summary, the present calculations of RbNbO$_3$ and RbTaO$_3$ properties and their
comparison with the properties of other ferroelectrics show that rubidium
niobate is an interesting ferroelectric material with high nonlinear optical
and electro-optical properties, and rubidium tantalate is an incipient
ferroelectric.

The calculations presented in this work were performed on the laboratory
computer cluster (16~cores).

This work was supported by the Russian Foundation for Basic Research,
project no. 13-02-00724.


%
\providecommand{\BIBYu}{Yu}

\end{document}